# Reinforcement-guided generative protein language models enable de novo design of highly diverse AAV capsids

Lucas Ferraz[1,*], Ana F. Rodrigues[1,*,✉], Pedro Giesteira Cotovio[1], Mafalda Ventura[3], Gabriela Silva[3], Ana Sofia Coroadinha[3], Miguel Machuqueiro[2], Catia Pesquita[1]

1 - LASIGE, Faculdade de Ciências da Universidade de Lisboa, Lisboa, Portugal

2 - BioISI, Faculdade de Ciências da Universidade de Lisboa, Lisboa, Portugal

3 - iBET – Instituto de Biologia Experimental e Tecnológica, Oeiras, Portugal

\* - Equal contribution

✉ - Correspondence should be addressed to A. F. Rodrigues (afdrodrigues@fc.ul.pt)

**Abstract**

Adeno-associated viral (AAV) vectors are widely used delivery platforms in gene therapy, and the design of improved capsids is key to expanding their therapeutic potential. A central challenge in AAV bioengineering, as in protein design more broadly, is the vast sequence design space relative to the scale of feasible experimental screening. Machine-guided generative approaches provide a powerful means of navigating this landscape and proposing novel protein sequences that satisfy functional constraints. Here, we develop a generative design framework based on protein language models and reinforcement learning to generate highly novel yet functionally plausible AAV capsids. A pretrained model was fine-tuned on experimentally validated capsid sequences to learn patterns associated with viability. Reinforcement learning was then used to guide sequence generation, with a reward function that jointly promoted predicted viability and sequence novelty, thereby enabling exploration beyond regions represented in the training data. Comparative analyses showed that fine-tuning alone produces sequences with high predicted viability but remains biased toward the training distribution, whereas reinforcement learining-guided generation reaches more distant regions of sequence space while maintaining high predicted viability. Finally, we propose a candidate selection strategy that integrates predicted viability, sequence novelty, and biophysical properties to prioritize variants for downstream evaluation. This work establishes a framework for the generative exploration of protein sequence space and advances the application of generative protein language models to AAV bioengineering.

## 1. Introduction

Adeno-associated viral (AAV) vectors are among the most widely used platforms for therapeutic cargo delivery in gene therapy. Their extensive clinical evaluation [1], sustained transgene expression [2], and design flexibility afforded by capsid [3] and genome engineering [4] have driven continued expansion into new therapeutic indications [5]. First-generation AAV vectors yielded major clinical successes and enabled early market-approved products, including Luxturna® and Zolgensma®. However, accumulated clinical experience has revealed the need for next-generation vectors with reduced immunogenicity, improved transduction efficiency, or lower organ toxicity [6]. Additionally, other properties remain active targets for optimization, including enhanced tissue specificity, improved manufacturability, and better control of transgene expression. Addressing these challenges requires the ability to efficiently explore large regions of capsid sequence space while preserving strict functional constraints.

Protein engineering is limited by the vast size of possible sequence space relative to the scale of feasible experimental screening. Even for relatively well-characterized proteins, only an infinitesimal fraction of possible sequence variants can be experimentally evaluated. Therefore, computational approaches capable of proposing novel, functionally plausible protein sequences have become increasingly important for guiding experimental efforts. In recent years, machine learning-guided protein design has emerged as a powerful strategy for navigating sequence space and accelerating the discovery of functional AAV variants [7–14].

Generative artificial intelligence (AI) represents a particularly promising approach for protein design. Generative models learn the underlying distribution of complex biological sequences and use this learned representation to produce new variants. Early work in generative protein modeling explored architectures such as generative adversarial networks (GANs) [15], variational autoencoders (VAEs) [16], and graph neural networks (GNNs) [17]. More recently, advances in large-scale protein language models (PLMs) [18–20], flow-based [21], and diffusion-based models [22,23] have substantially improved generative models' ability to capture protein sequence patterns and structural constraints. These developments have enabled the generation of diverse proteins that maintain plausible structural and evolutionary characteristics.

In the context of AAV bioengineering, most generative studies have relied on VAEs [24–26], which learn a continuous latent representation of training sequences and enable the generation of new variants through sampling or interpolation. However, VAE-generated sequences often remain strongly biased toward the training data, and these models are susceptible to latent space collapse (e.g. [27]), in which the learned representation fails to capture meaningful sequence variability. As a result, such approaches have limited ability to explore distant regions of sequence design space that could contain highly novel variants. Accessing more distant regions of sequence space is important because it increases the likelihood of discovering variants with properties absent from existing capsids, enabling the identification of vectors with altered antigenicity, improved stability, or modified tissue tropism, thus expanding the functional repertoire available for gene therapy applications. Developing generative strategies that can effectively explore such novelty while preserving functional constraints remains a challenge.

PLMs are leading approaches for modeling protein sequence distributions and generating new variants. Pretrained on millions to billions of protein sequences, PLMs capture rich statistical patterns, long-range sequence dependencies, and evolutionary constraints encoded in natural proteins [28]. This large-scale pretraining provides a biologically informed prior that enables the generation of plausible sequences by meeting the *protein grammar* rules [29]. Additionally, PLM-based approaches offer greater scalability, easier training, and broader coverage of sequence space compared to other generative approaches such as GANs, GNNs, or diffusion-based models [30], making them attractive for advancing generative AAV capsid design. However, pretrained PLMs are inherently general-purpose models and do not explicitly encode the functional requirements of any particular bioengineering task. Consequently, they cannot reliably generate sequences optimized for specific functional properties, such as, in the case of AAVs, viability, manufacturability, or immune evasion. Addressing this limitation requires adapting the model to task-specific data through fine-tuning, a form of transfer learning that enables it to retain knowledge acquired during large-scale pretraining while incorporating functional signals relevant to the bioengineering objective [31]. This approach is particularly effective when experimental datasets exist or can be generated that explicitly capture the functional properties of interest, even if such datasets are small, which is common in practice.

After a model is fine-tuned to satisfy task-specific functional constraints, generating truly novel sequences remains a challenge. Protein sequence space is vast, and most combinations of residues are non-functional, meaning that naive sampling, even from a fine-tuned model, tends to produce sequences that are close to the training data. Achieving higher novelty requires actively guiding the model to explore regions of sequence space that are distant from known sequences. Reinforcement learning [32] provides a powerful framework for this goal by defining reward functions that simultaneously capture functional constraints and sequence novelty.

In this work, we develop a generative protein design framework based on fine-tuned PLMs and reinforcement learning to generate highly novel but functional proteins. We demonstrate this approach in the context of AAV2 capsid engineering which, represents a pioneer application of PLM-based generative modeling for AAV design. We first fine-tune a pretrained PLM on a large corpus of experimentally validated AAV2 capsid sequences to capture sequence patterns associated with the functional constraint of capsid viability. We then introduce a reinforcement learning-based training scheme guide generation toward variants predicted to retain capsid viability while actively promoting sequence novelty beyond that found in training data. In contrast to conventional fine-tuning approaches that rely solely on positive examples, this strategy explicitly teaches the model not only what constitutes viable sequences but also which regions of sequence design space to avoid [33,34], thus increasing the generation of functional variants. Finally, we propose a candidate selection strategy that integrates predicted viability, sequence novelty, and coverage of biophysical properties to prioritize the most promising variants for downstream experimental evaluation. This combination of generation, reinforcement learning-guided exploration, and candidate selection enables systematic discovery of diverse, functionally plausible protein sequences and establishes a generalizable workflow for PLM-based generative protein design that can be applied beyond AAV capsid design.

## 2. Results

### 2.1. Training and sequence generation

To evaluate the potential of PLM fine-tuning for generative AAV capsid design, we used ProGen [30, 31], a generative PLM pre-trained on hundreds of millions of diverse protein sequences spanning a wide range of families and functions. While more recent PLMs such as ESM3[18] provide powerful capabilities for protein generation, their large-scale and multimodal design makes iterative reinforcement learning optimization computationally demanding. We therefore employed the ProGen architecture as a tractable generative backbone that enables direct integration of fine-tuning and sequence-level reinforcement learning. Because our goal was not to benchmark different PLMs but to evaluate how generative PLMs can be adapted for functional AAV capsid design, ProGen provided a practical balance between generative capability, computational tractability, and flexibility for reinforcement learning–based optimization.

The pre-trained model was fine-tuned using either a classical fine-tuning approach or fine-tuning followed by reinforcement learning training, leveraging the AAV2 capsid viability dataset from Bryant *et al.* (2021) [11], with the model prior to any fine-tuning or retraining, being evaluated as a baseline (**Fig. 1a**). In the classical fine-tuning approach, only sequences labeled as viable (positive sequences) were used for training (**Fig. 1b**). This fine-tuned model then served as the initialization for the reinforcement learning strategy (**Fig. 1c**). The reinforcement learning architecture was designed to balance sequence novelty with functional constraints through two complementary branches. The diversity branch uses embeddings from ESM2 model [37] which were kept frozen to provide general-purpose sequence representations unbiased with respect to capsid viability signals. Here, novelty is quantified by comparing each generated sequence to the reference sequence using the cosine distance between embeddings. Because these embeddings capture the overall structural and evolutionary context of a protein, the cosine distance between them reflects how different a new sequence is from the reference sequence in a biologically informed latent space, rather than merely raw sequence identity. The distribution of cosine distance values of sequences in the fine-tuned generated set is used to express novelty scores as percentiles, scaled such that a value of 1 corresponds to the maximum novelty observed in this set. In the reward function, novelty contributes multiplicatively such that higher novelty leads to a larger reward contribution. For sequences exceeding the maximum observed novelty, this contribution is further amplified through power-law scaling, effectively creating a tail that incentivizes exploration beyond the training distribution. The functional branch uses ProtBERT [38] with a classification head, fine-tuned on experimental data of AAV2 capsid viability. ProtBERT provides a sequence-level representation equivalent in capability to ESM2, but using a separate model for the functional branch ensures that novelty and viability are treated independently.

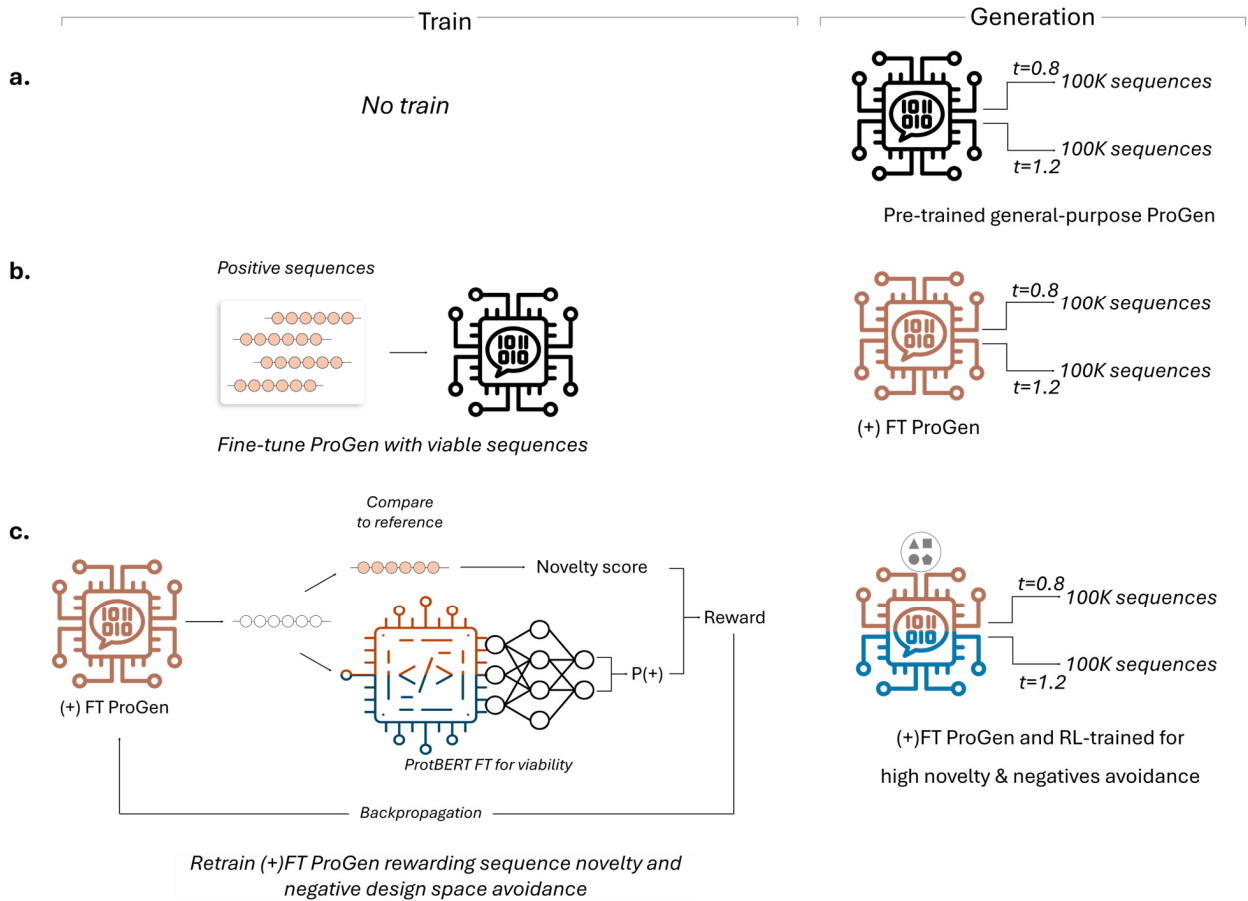

**Figure 1 | Overview of training and sequence generation strategies**. The pre-trained, general-purpose ProGen model was first used as a baseline to generate new sequences without additional training **(a)**. Next, ProGen was fine-tuned using only positive (viable) sequences **(b)**. This fine-tuned model ((+)FT ProGen) was further trained using a reinforcement learning (RL) scheme that simultaneously rewards sequence novelty relative to the reference sequence and penalizes exploration of the negative sequence design space **(c)**. This figure was created using icons from The Noun Project, used under the CC BY 3.0 license: *language model* by ari supriharyati, *Embedding* by Vectors Point, and *Neural Network* by karyative.

With each of the three model variants (pre-trained, fine-tuned, and fine-tuned plus reinforcement learning) and using intermediate (0.8) or high (1.2) sampling temperatures, 100000 sequences were generated (**Fig. 1, Table 1**).

**Table 1 | Intersection of unique sequences across strategies and the training dataset**

|  |  | Train set | Pre-trained ProGen | | (+) FT ProGen | | (+) FT ProGen & RL | |
|---|---|---|---|---|---|---|---|---|
|  |  |  | t = 0.8 | t = 1.2 | t = 0.8 | t = 1.2 | t = 0.8 | t = 1.2 |
| **Train set** |  | 293305 | 0 | 0 | 161 | 37 | 0 | 6 |
| **Pre-trained ProGen** | **t = 0.8** |  | 100000 (0% viable) | 2 | 0 | 0 | 0 | 0 |
|  | **t = 1.2** |  |  | 100000 (0% viable) | 0 | 0 | 0 | 0 |
| **(+) FT ProGen** | **t = 0.8** |  |  |  | 82031 (98% viable) | 75 | 0 | 0 |
|  | **t = 1.2** |  |  |  |  | 72509 (91% viable) | 0 | 0 |
| **(+) FT & RL ProGen** | **t = 0.8** |  |  |  |  |  | 330 (100% viable) | 317 |
|  | **t = 1.2** |  |  |  |  |  |  | 9403 (99% viable) |

Except for the train set, the diagonal shows the number of unique sequences (out of the 100000) generated in each case. Percentages along the diagonal values indicate the proportion of unique sequences in each case that were classified as positive (viable) by the classifier described in Rodrigues *et al.* (2026) [31].

The general-purpose pre-trained model produced exclusively unique sequences within each batch, with only 2 sequences shared between the 0.8 and 1.2 temperature conditions (**Table 1**). None of the generated sequences overlapped with the fine-tuning dataset, showing that the pre-trained model captures general protein grammar but lacks functional guidance resulting in all sequences being classified as non-viable. Fine-tuned models also generated a high proportion of unique sequences (82% at t=0.8 and 73% at t=1.2) with 75 sequences shared between the two temperature conditions. A limited overlap with the training set was observed (161 and 37 sequences at temperatures 0.8 and 1.2, respectively), consistent with a residual bias toward the training data (**Table 1**). Nonetheless, the fraction of new and unique sequences remained very high. Among these unique sequences, 98% (for t=0.8) and 91% (for t=1.2) were classified as viable by an AAV2 viability classifier described previously [31]. By contrast, the fine-tuned reinforcement learning model generated fewer unique sequences, 330 (t=0.8) and 9403 (t=1.2), among the 100000 sequences generated per temperature, the majority classified as viable (100% for t=0.8, and 99% for t=1.2). As with the classical fine-tuning approach, some sequences were shared across temperatures, and a small fraction overlapped with the training set. This reduced number of unique sequences reflects the selective pressure imposed by the reward, which simultaneously promotes functional viability and sequence novelty while discouraging non-functional regions of sequence space.

We further analysed the mutational landscapes of the sequences generated by each model and compared them with those of the training dataset, comprising both positive and negative sequences (**Fig. 2**).

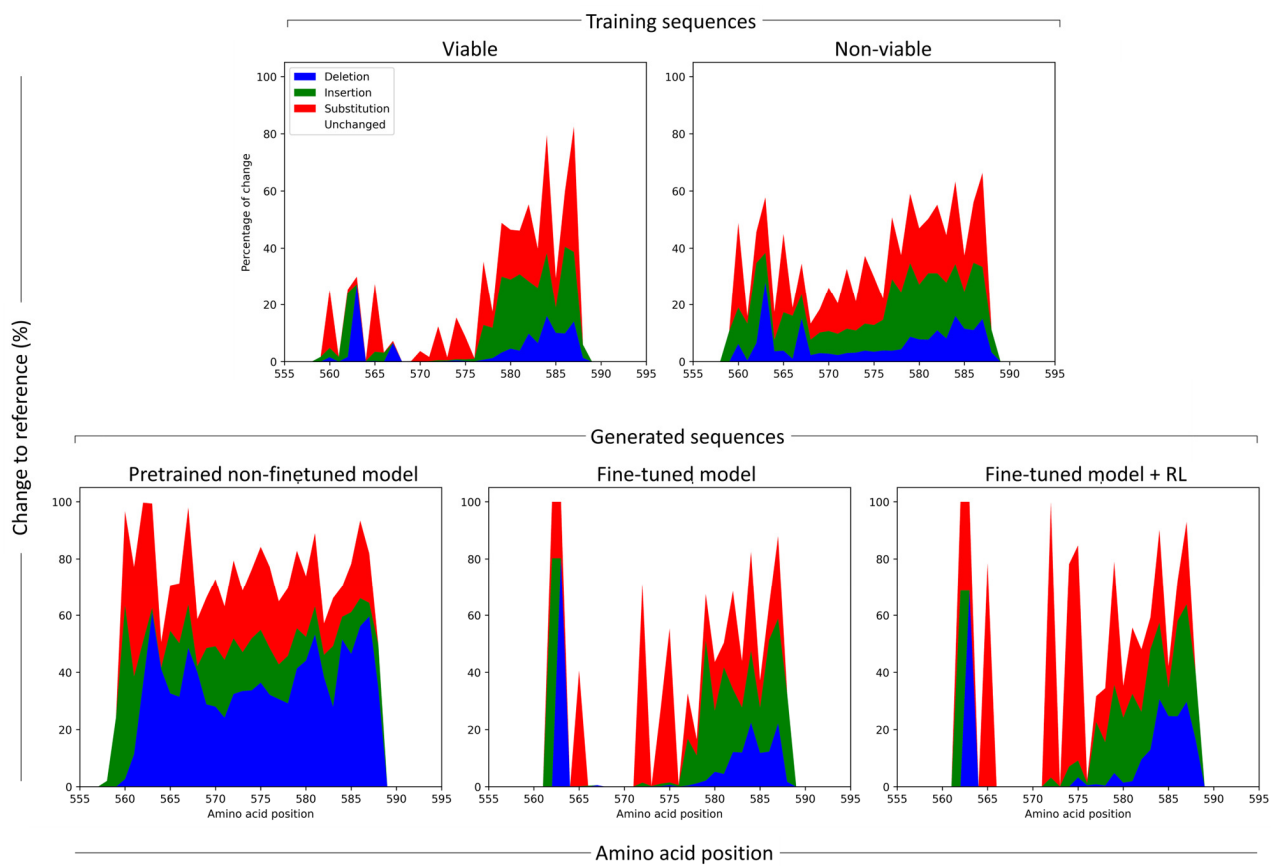

**Figure 2 | Mutational landscape analysis.** Distribution of mutation types (deletions, insertions, and substitutions) within the targeted region (amino acid positions 561-588), shown as the percentage of change relative to the reference sequence. Data are presented for positive (viable) and negative (non-viable) sequences for the training set, sequences generated by the pre-trained ProGen model without fine-tuning, sequences generated by the ProGen model fine-tuned on positive (viable) sequences and sequences generated by the fine-tuned ProGen model further optimized with reinforcement learning (RL). Generated datasets include sequences produced with sampling temperatures $t = 0.8$ and $t = 1.2$; only unique sequences are included within each group.

The mutational landscape of sequences generated by the fine-tuned models closely resembled that of the positive training set, recapitulating a previously identified viability signature [31], where viable sequences preferentially avoid mutations within the region spanning amino acids 567-576. When mutations occur in this region, they are predominantly substitutions rather than insertions or deletions. This region corresponds to a segment of the protein that is more buried within the capsid structure and is therefore likely critical for maintaining capsid integrity, providing a plausible structural basis for the observed mutation avoidance (**Fig. 3**). Sequences generated by the reinforcement-guided model not only preserve this viability signature but also exhibit higher-intensity mutational targeting in the permitted regions outside the restricted segment, revealing that this model actively drives exploration of novel sequence variants by concentrating mutations in regions compatible with viability. On the other hand, sequences generated by the general-purpose pre-trained model exhibited a markedly less structured mutational landscape, resembling neither the positive nor the negative sequences from the set used for fine-tuning. Instead, mutations were distributed broadly across the 560-588 region, with high-intensity targeting throughout. This, and especially the disruption of the viability signature, likely led to the classification of all of these sequences as non-viable.

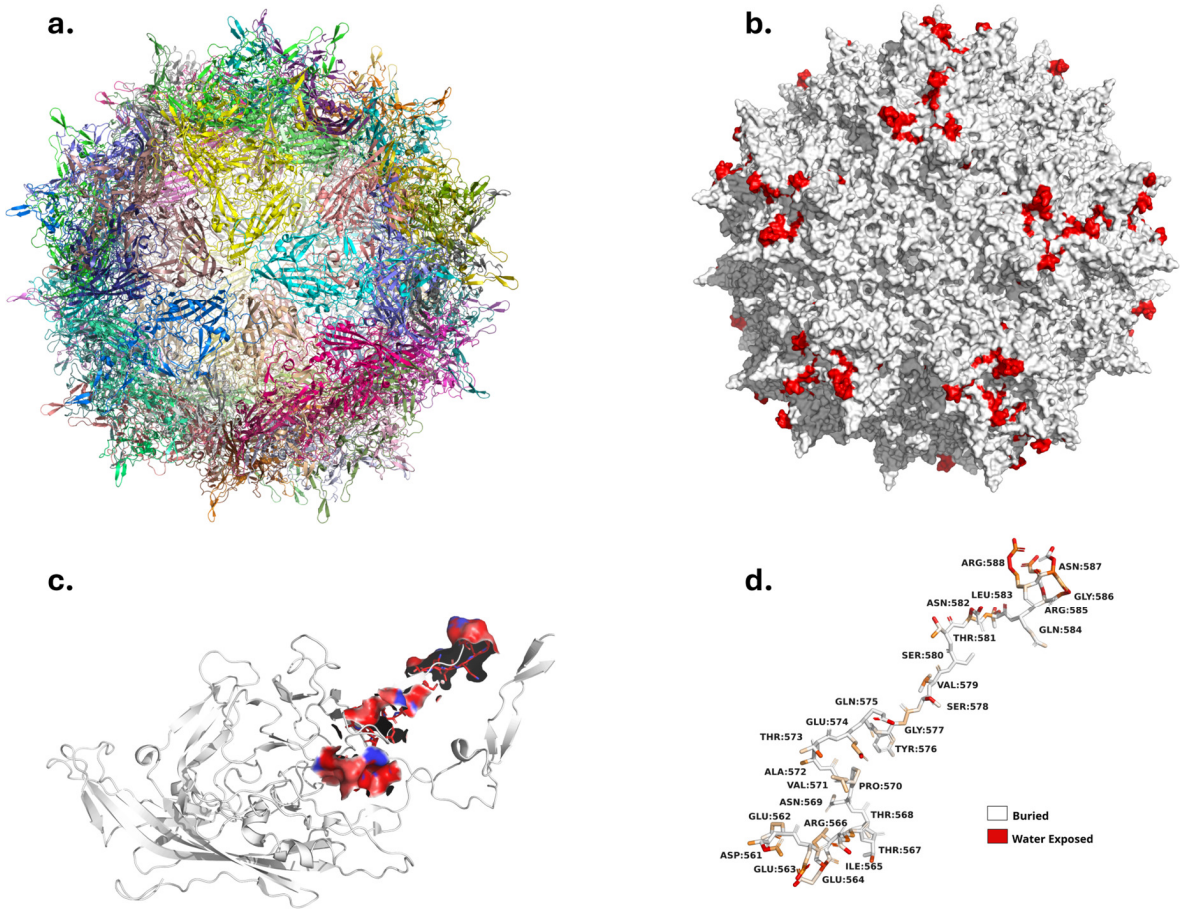

**Figure 3 | 3D structure of the adeno-associated virus 2 (AAV2).** Cryo-EM structure of the AAV2 rep-capsid packaging complex containing a 60-subunit icosahedral protein shell structure (PDB ID: 8FYW [39]) (**a**). The 561-588 region (colored in red) of the single-chain protein (735 aa) is partially exposed to solvent (**b** and **c**). Residues in the 567-576 range are among the most buried in the studied segment (**d**).

## 2.2. Evaluation of sequence novelty

Unique sequences from each design strategy were analyzed using embedding-based sequence representations (**Fig. 4**). Global sequence ESM2 embeddings were used as a sequence representation. The distribution of sequences in embedding space was visualized with t-distributed Stochastic Neighbor Embedding (t-SNE) and colored in three complementary ways: i) by sequence source (**Fig. 4a**), ii) by novelty scores (**Fig. 4b**), and iii) by predicted viability (**Fig. 4c**). The distribution of novelty scores was also summarized using violin plots (**Fig. 4d**).

Sequences generated by the pre-trained general-purpose model were found to occupy a region of the sequence space distinct from both the training set and the fine-tuned sequences (**Fig. 4a**). However, this did not correspond to very high novelties (**Fig. 4b**) and, as seen before, none of these sequences were predicted to be viable (**Fig. 4c**). Then, training sequences, fine-tuned generated sequences, and reinforcement learning generated sequences laid in different regions of sequence space, forming two main clusters displaying a gradient pattern (**Fig. 4a**): the upper-right region is enriched for training sequences, the central region contains a mix of fine-tuned and reinforcement learning sequences, and the left-most region is dominated by reinforcement learning sequences. While no cluster is entirely pure, the predominance of each generation strategy along this gradient is evident, reflecting how reinforcement learning systematically shifts sequences toward previously unexplored regions of sequence space. This gradient also corresponded to an increase in sequence novelty, with the highest values observed among the left-most reinforcement learning-generated sequences (**Fig. 4b**). Importantly, this exploration of novel sequence regions did not sacrifice predicted viability (**Fig. 4c**), showing that our reinforcement learning strategy expands sequence diversity while maintaining functional plausibility. Reinforcement learning-generated sequences also achieved the highest novelty scores overall, exceeding both the maximum values observed in fine-tuned sequences and the training set (**Fig. 4d**), demonstrating the effectiveness of our approach in driving functional exploration beyond the limits of the original dataset.

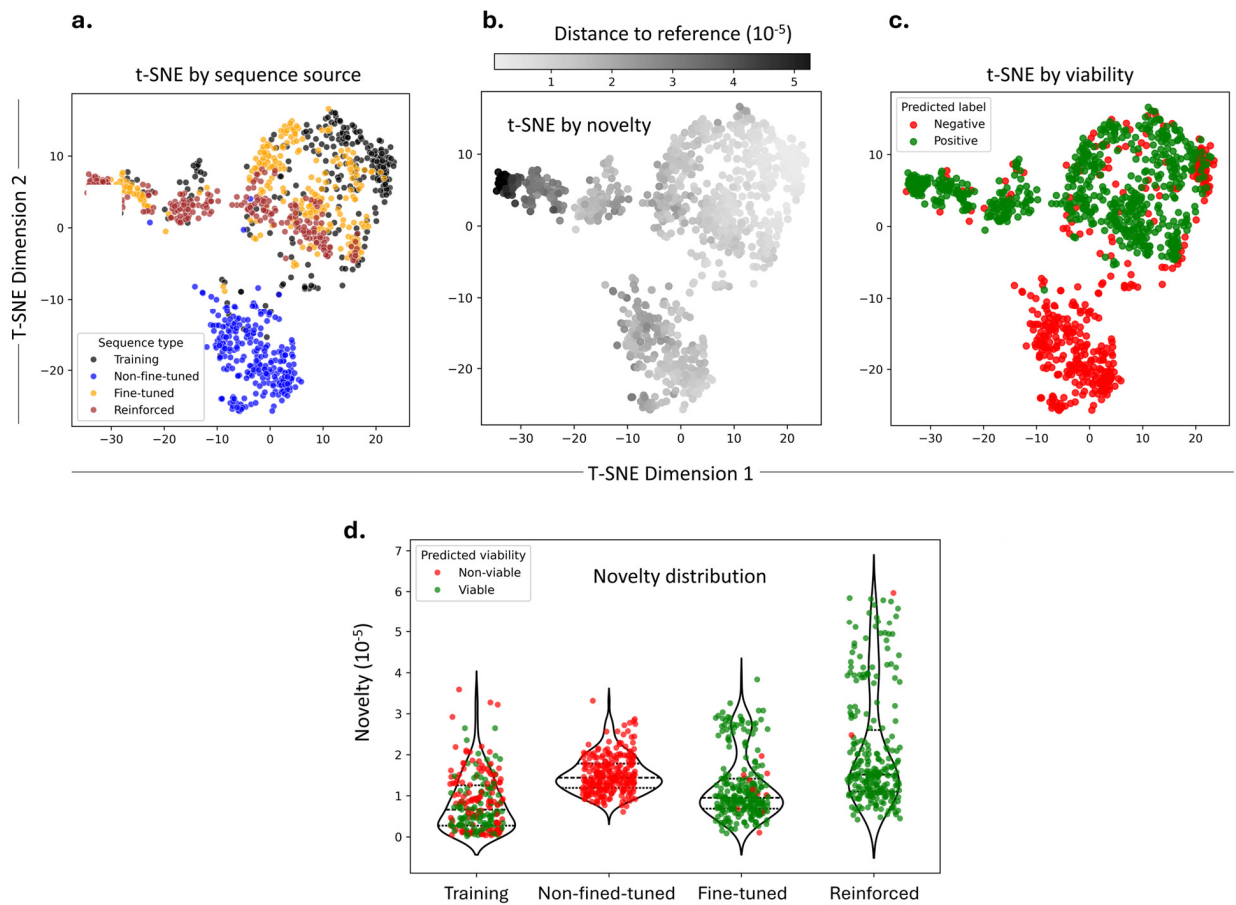

**Figure 4 | Sequence embedding-based analysis of novelty and viability.** Unique sequences from each design strategy were embedded using ESM2 representations and projected into two dimensions with t-SNE. Panels show sequences colored by design strategy (**a**), novelty relative to the reference sequence (cosine distance, **b**), and predicted viability from an ESM2-based classifier (**c**). Panel (**d**) summarizes the distribution of novelty scores with violin plots. In all panels, a random sample of 500 sequences is shown.

### 2.3. Evaluation of biophysical properties

Focusing on the reinforced-source sequences, where higher novelty was achieved, we sought a strategy to select candidates for experimental validation. Using predicted viability alone was insufficient because, since most sequences generated by this model were predicted as viable. To further narrow the set and improve candidate selection for laboratory testing, we analyzed two biophysical properties known to influence protein stability and functional maintenance [40]: polarity and charge (**Fig. 5a**). These properties serve as developability filters, highlighting sequences with favorable folding and solubility profiles and providing additional criteria for selecting candidates from the high-viability, high-novelty set. The analysis focused on the 28-residue region window subjected to change corresponding to positions 561-588 in the reference sequence, which can extend to a 35-residue window in variants with insertions.

Positive sequences were found to display slightly higher polarity than negative sequences, indicating a modest enrichment of hydrophilic residues and suggesting that higher polarity in this region is associated with viability and developability. Fine-tuned and reinforced sequences accentuated this trend, shifting the distribution toward increased polarity. For net charge, negative sequences tended toward neutrality, whereas non-fine-tuned sequences were shifted further toward positive values. Positive (viable), fine-tuned, and reinforced sequences, in contrast, were slightly biased toward anionicity, indicating that excessive positive charge in this region is detrimental. Despite some identifiable trends, no drastic differences between positive and negative sequences were found that would allow defining strict cut-offs for sequence selection. Therefore, we adopted a grid-based sampling strategy along the net polarity and net charge axes (**Fig. 5b**). In this approach, sequences were sampled evenly across the 2D space defined by these two axes, while avoiding extreme values. To achieve this, we restricted both axes to the middle 90% of the positive sequence values, i.e., excluding the bottom and top 5% values. Within these boundaries, lied more the more the 7000 reinforced-source sequences with p(viable)>0.5. This bounded 2D space was therefore considered a representative landscape of diverse biophysical parameter combinations from which to sample novel sequences. To systematically explore this space, it was partitioned into discrete regions (9 and 25 bin combinations), illustrating both coarse- and fine-grained sampling strategies. Each region corresponds to a distinct combination of net polarity and net charge, capturing different areas of biophysical diversity. Increasing the number of regions enables a finer resolution of this landscape and a more comprehensive exploration of viable sequence space. Within each region, we selected the sequence with the highest novelty score (**Figure 5b** and **Table 2**), ensuring that the final selection not only maintains high predicted viability but also maximizes biophysical and sequence novelty.

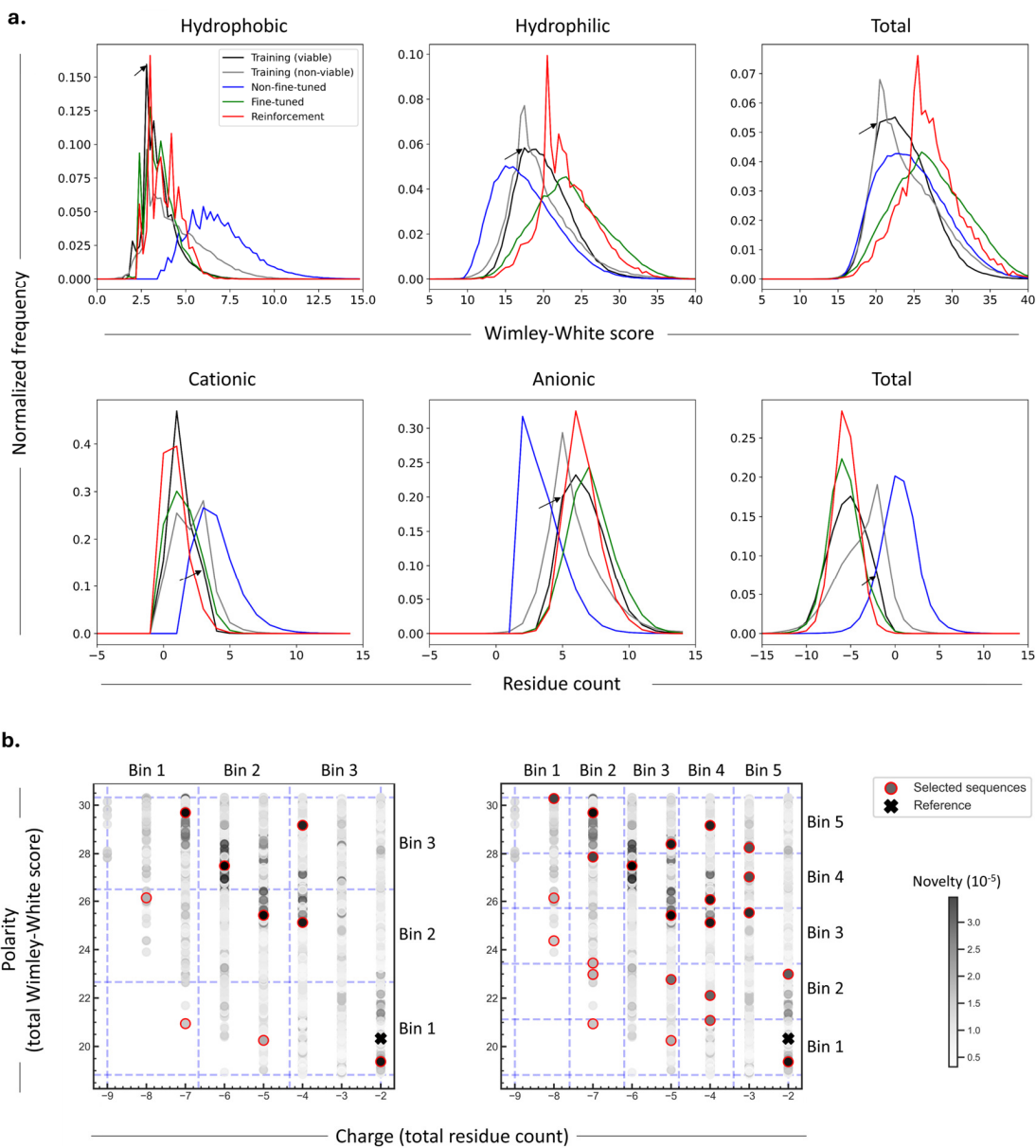

**Figure 5 | Polarity- and charge-based analysis and strategy for candidate sequences selection.** Top panels show the distribution of hydrophobic, hydrophilic, and total polarity, calculated using the Wimley-White score (see Methods), and bottom panels show the distribution of cationic, anionic, and total charge (**a**). Distributions are shown for the training sequences (positives and negatives) and sequences generated by the non-fine-tuned, fine-tuned, and fine-tuned plus reinforcement learning models. Reinforced learning-source sequences classified as positive with high confidence (P(positive)>0.5) are plotted across the total polarity and total charge axes, colored by their novelty score (**b**). The selection space was divided into nine (left panel) or 25 (right panel) sub-regions (defined by a given polarity and charge bin intersection) and within each sub-region, the sequence with the highest novelty score presents as the choice candidate for downstream evaluation. The reference sequence is indicated by an arrow in the distribution plots and by a cross in the selection grid**.**

**Table 2 | Candidates for further downstream analysis and/or experimental validation (25-region search)**

| Fragment* | Polarity bin | Charge bin | Novelty score ($10^{-5}$) |
|---|---|---|---|
| DEEEIRTTNPVATEQYGSVSTNLQRGNR** | 1 | 5 | - |
| DECEIATTNPVAYECWGSVCFTWENNQDE | 1 | 2 | 2.44 |
| DESCIATTNPVAYEGWGCVCIENDSMTTG | 1 | 3 | 3.04 |
| DESCIRTTNPVAYECWGCCACYNWETSDGGI | 1 | 5 | 4.20 |
| DESCIRTTNPVAYECWGCCCWDCNPHTTGFNIC | 1 | 5 | 6.40 |
| DECEIATTNPVAYECWGSVCSTWENDMQGEG | 2 | 2 | 2.91 |
| DESCIRTTNPVAYECWGCCYNCAPTPEGPNMDD | 2 | 3 | 4.40 |
| DESCIRTTNPVAYECWGCCQSCSNWDALTNIEGNA | 2 | 5 | 4.74 |
| DESCIRTTNPVAYECWGCVCSTWVNMQDTRGGANYQAFC | 2 | 5 | 4.92 |
| DECEIATTNPVAYEQWGCVCDNDLYDQD | 2 | 1 | 2.45 |
| DECEIATTNPVAYECWGSVCSTWEQGDTNDG | 3 | 2 | 2.86 |
| DEACISTTNPVAYEGYGQCQCCSMWGEHNAEFFNSDTQLSCCC | 3 | 3 | 6.67 |
| DEACISTTNPVAYEGYGQCQCCSMWGEHNAEFFNSDTQLHQCC | 3 | 5 | 6.22 |
| DESCIRTTNPVAYECWGCVCSTWNEDHQTGNSDNRMYFCC | 3 | 5 | 5.87 |
| DECEIATTNPVAYEQWGCVCLDDDQMLNMND | 3 | 1 | 2.99 |
| DEACISTTNPVAYEGYGQCQCCSMWGEHNAEFFNSDTQLEEC | 4 | 2 | 5.41 |
| DEACISTTNPVAYEGYGQCQCCSMWGEHNAEFFNSDTQLDCCC | 4 | 3 | 6.85 |
| DEACISTTNPVAYEGYGQCQCCSMWGEHNAEFFNSDTQLHGCC | 4 | 5 | 6.17 |
| DESCIRTTNPVAYEGWGQCVCSMWGEHNAEFFNSDTQLTYCCR | 4 | 5 | 5.22 |
| DEACISTTNPVAYEGYGQCQCCSMWGEHNAEFFNSDTQLEDCE | 4 | 1 | 5.48 |
| DEACISTTNPVAYEGYGQCQCCSMWGEHNAEFFNSDTQLDCCD | 5 | 2 | 6.55 |
| DESCIRTTNPVAYEGWGQCQCCSMWGEHNAEFFNSDTQLMCCD | 5 | 3 | 6.24 |
| DESCIRTTNPVAYECWGCCQACPNWGEDHNAEFFISDTQPNKC | 5 | 5 | 6.06 |
| DESCIRTTNPVAYECWGCCACWCNEFSQDPGTKTSYLAKAD | 5 | 5 | 5.13 |

\* Region corresponding to the 561-588 in the reference sequence. ** Reference sequence

## 3. Discussion

AAV capsid design is an important and rapidly advancing area of research, enabling the development of vector variants with improved therapeutic properties, yet, as in protein bioengineering more broadly, it is challenged by the need to navigate vast sequence spaces while preserving strict functional constraints. Generative PLMs offer a powerful means to address this challenge, providing advantages over other generative strategies, including scalability, ease of training, and a biologically informed ability to produce structurally plausible protein sequences [28,29]. However, the extent to which these models can generate functionally meaningful AAV capsid variants, and how different training strategies influence this capability remain largely unexplored. In this work, we address this question by evaluating three model configurations: a general-purpose pre-trained PLM as a baseline, a task-adapted model obtained through classical fine-tuning, and a reinforcement learning-guided version designed to balance functional optimization with sequence novelty.

The comparative analysis across the three model types (**Table 1**) highlighted their distinct behaviors. Classical fine-tuning generated many unique sequences that largely retained predicted viability (91-98%), with an expected residual bias toward the training data. In contrast, the reinforcement learning-guided model produced fewer unique sequences, reflecting the trade-off between maintaining viability and exploring sequence novelty beyond the training set. A small bias toward the training data persisted, likely because the model started from the fine-tuned baseline, but the overall predicted viability of the reinforcement learning strategy was higher (99-100%), demonstrating its ability to actively avoid non-functional regions of the sequence design space. While additional unique sequences could be obtained by sampling from multiple random seeds rather than generating 100000 sequences in a single batch, we retained a single-seed strategy to ensure a fair comparison across models. Notably, reinforcement learning required initialization from the fine-tuned model, as starting from the pre-trained, non-fine-tuned model failed to converge. This is because sequences generated by the pre-trained model were almost always classified as non-viable (from the 100000 sequences generated here, all were negative), resulting in little reward and preventing effective learning. These results highlight a known limitation of general-purpose PLMs, since without task-specific adaptation, they cannot reliably generate sequences that meet functional criteria. This limitation is not specific to AAV capsids but applies broadly in protein engineering, where most of the sequence space is non-functional. Consequently, starting with a functionally biased prior is essential for successful reinforcement learning-guided sequence generation.

The mutational landscape analysis provided further insights into the biological constraints of sequence design and revealed how different generative strategies balance functional conservation with sequence exploration (**Fig. 2**). Fine-tuned models effectively reproduced a critical viability signature, demonstrating that task-specific adaptation allows the model to respect structural constraints necessary for protein function. Reinforcement learning went further, actively concentrating mutations in tolerated regions while actively avoiding deleterious positions. This high-intensity mutational targeting within permissible regions formed the basis for generating sequence more novel than anything observed in the training set, driving functional exploration beyond the highest diversity previously reported for AAV2 [11]. This enhanced novelty facilitates the discovery of variants with altered properties, including modified stability, antigenicity, or tissue specificity, which is essential for future applications of AAV capsid design. In contrast, sequences produced by the general pre-trained model lacked such constraints, showing broad, unstructured mutational patterns that disrupt viability signatures. These sequences were predicted to be non-viable and would likely fail experimental validation, highlighting the limitations of unguided generative approaches.

Diving deeper into the diversity and novelty of the generated sequences, t-SNE analysis revealed a gradual shift in space occupied by the different sequences, from training, to fine-tuned, and then to reinforcement learning-generated sequences (**Fig. 4**). Fine-tuned sequences clustered close to the training set, as expected since the model was explicitly trained on positive sequences, recapitulating their key functional patterns. Reinforcement learning-generated sequences occupy regions that were increasingly distant from the training data, reflecting the influence of the novelty-driven reward. Some proximity to the fine-tuned sequences remains, consistent with the reinforcement learning model being initialized from the fine-tuned model. Notably, the most isolated reinforcement learning-generated cluster corresponded to the sequences with the highest novelty scores, providing strong evidence that the novelty metric successfully drives exploration into previously unobserved sequence variants and aligns with broader patterns of global sequence diversity captured by t-SNE. At the same time, the most distant cluster in the t-SNE space was formed by sequences generated from the non-fine-tuned pretrained model, which exhibit only average novelty values. This apparent discrepancy arises because t-SNE captures many dimensions of sequence variation, whereas the novelty score measures divergence from a reference sequence within a biologically informed latent space. Consequently, while highly novel sequences tend to occupy distinct regions of t-SNE space, occupying a distant t-SNE region does not necessarily indicate high novelty. An important consideration is the interpretation of the novelty score used to guide reinforcement learning. In this work, novelty is computed as the distance between sequence embeddings generated using pretrained ESM2 model. Because these embeddings encode rich information about protein sequence, structure, and evolutionary relationships, distances in this space provide a biologically informed measure of divergence from known capsid sequences. This metric is therefore a good proxy for exploring regions of sequence space that are underrepresented in the training data.

Selecting candidates for experimental validation from the reinforcement learning-generated sequences proved challenging. Because the reinforcement learning reward penalizes non-viable sequences, the model predominantly produced sequences predicted to be viable, limiting the utility of predicted viability as a discriminative filter. Moreover, the viability classifiers used herein were trained on the same experimental data as that used for model fine-tuning, and are therefore biased toward recognizing viability within the sequence space sampled during training, making them prone to misclassifying highly divergent, novel sequences, particularly through false negatives. While this limitation could be addressed through alternative strategies such as retraining on expanded datasets or integrating orthogonal functional assays, we did not attempt to solve it here. Instead, our approach applied a positive-filtering step based on predicted viability, meaning that some genuinely functional sequences may have been excluded due to their novelty, and was complemented by a selection strategy that, although informed by the training data, does not rely on it directly. This strategy explores the space of two biophysical parameters, polarity and charge, and selects sequences with high novelty within distinct regions of this combined parameter space (**Fig. 5**). This strategy leverages established principles of protein developability, as intermediate polarity and charge values are generally associated with proper folding, solubility, and functional maintenance, whereas extreme values are more likely to compromise these features [40]. In practice, such extremes were avoided by excluding the 5% tails observed in the distribution of viable sequences. While the selected candidates are still indirectly informed by the training dataset, the grid-based biophysical framework provides an orthogonal axis of exploration, broadening the search space in a controlled and interpretable manner. The number of candidates could be readily increased subdividing the grid into more regions, enabling finer-grained exploration of the polarity-charge landscape. At the same time, the number of selected candidates (**Table 2**) represents a manageable set for downstream analyses, such as molecular dynamics simulations or other computational biophysical evaluations, which are computationally intensive and therefore most appropriately applied at later stages, when the candidate pool has already been substantially reduced. It is also manageable for direct experimental evaluation.

This work demonstrates the potential of generative protein language models for AAV capsid design. It shows that general-purpose PLMs alone are insufficient to generate functional sequences, but task-specific fine-tuning enables the recovery of key functional constraints (here, viability), while reinforcement learning drives exploration toward highly novel variants. We further introduce a biophysically informed candidate selection framework that provides a straightforward strategy for identifying diverse variants for downstream evaluation, based on criteria orthogonal to those used during model training. Beyond AAV engineering, this work establishes a robust framework that integrates generative PLMs, reinforcement learning, and biophysics-based selection to help navigate vast protein sequence spaces in bioengineering studies.

## 4. Materials and methods

### 4.1. Dataset and data pre-processing

The dataset used in this work is derived from the AAV2 capsid viability dataset reported by Bryant *et al.* (2021) [11]. Sequences in this dataset contain changes exclusively in the region corresponding to amino acids 561-588 in the reference sequence, comprising single and multiple mutations, including substitutions, deletions, and insertions, either alone or in combination. This region is frequently targeted in AAV bioengineering because it plays a key role in determining the vector's tropism, immune response, and overall stability [41], making it a prime candidate for modifications to optimize therapeutic potential. Data was processed as described previously [31]. Briefly, mutated capsid fragments were used to reconstruct full-length AAV2 capsid protein sequences, sequences were standardized in format, and duplicate or redundant entries were removed. After preprocessing and filtering, the final dataset comprised 293835 unique capsid sequences experimentally validated for viability. From these, the set of positive sequences was used for fine-tuning the generative model.

### 4.2. Mutation landscape analysis

Mutation landscape analysis followed the procedure described in Rodrigues *et al.* (2026) [31]. Briefly, each sequence in the dataset was compared to the reference sequence (NCBI accession P03135.2) to identify substitutions, deletions, and insertions using the *pairwise2* module from Biopython. The frequency and type of mutations at each sequence position (mutation landscape) were then quantified using custom functions as described previously [31].

### 4.3. Sequence evaluation parameters

*Sequence viability*

Generated sequences were evaluated for predicted viability using a previously developed model for AAV2 viability classification described by Rodrigues *et al.* (2026) [31]. This classifier consists of a pretrained PLM (ESM2 [37]) augmented with a linear classification head for binary prediction. The architecture was trained end-to-end on experimentally validated AAV2 sequences from the viability dataset reported by Bryant *et al.* (2021) [11], enabling the classification signal to propagate through the encoder and optimize sequence representations for viability prediction. For inference, predictions were based on the global sequence representation derived from the CLS (classification) token, which demonstrated the strongest predictive performance for this viability-tuned classifier [31].

*Sequence novelty*

To quantify sequence novelty, sequence similarity was evaluated against the reference AAV2 capsid. For each sequence, global embeddings based on the CLS token were generated using the pretrained ESM2 model. The reference AAV2 sequence was embedded in the same manner. Cosine distances between the embeddings of each sequence and the reference were calculated, providing a continuous measure of divergence from the reference capsid and serving as a proxy for sequence novelty.

*Sequence polarity and charge*

Polarity and charge were computed for the sequence region corresponding to positions 561-588 of the AAV2 capsid reference sequence, which encompasses the mutationally diversified region used during sequence design and can extend in length in some sequences when insertions are present. Polarity was quantified using the Wimley–White index, calculated by multiplying the residue solvent-accessible surface area by the Wimley–White hydrophobicity scale, based on experimentally derived values for each amino acid based on its energetic preference for membrane interfaces [42]. For each extracted window, hydrophobic and hydrophilic contributions were calculated separately by summing the magnitudes of negative and positive Wimley-White values, respectively. The total polarity score was defined as the sum of these two contributions, providing a measure of the overall polarity magnitude within the sequence segment. Charge was computed withing the same regions by counting positively and negatively charged residues and assuming their most common protonation states at neutral pH. The cationic component corresponded to the number of arginine (R), and lysine (K) residues, while the anionic component corresponded to the number of aspartate (D) and glutamate (E) residues. Net charge was calculated as the difference between these two components, reflecting the electrostatic balance of the sequence segment.

### 4.4. Supervised fine-tuning

Supervised fine-tuning was performed on the open-source pretrained ProGen2-small model (https://huggingface.co/hugohrban/progen2-small) using viable AAV2 sequences. Each sequence in the dataset consists of two invariant regions, the prefix (positions 1-560 of the reference sequence) and the suffix (last 147 positions of the reference sequence), flanking a mutationally diversified fragment (positions 561-588 in the reference sequence, totalizing a 28-amino acid window, which can extend to 35 residues in some variants due to insertions). Because only the central fragment varies across sequences, the fine-tuning loss was computed exclusively on this variable region, while the prefix and suffix were masked to prevent the model from overfitting to invariant positions. Formally, the loss function was defined as:

Eq. (1):

$$L = -\frac{1}{\sum_t m_t} \sum_t m_t \log p_\theta(x_t \mid x_{<t})$$

where $x_t$ is the amino acid at position $t$, $p_\theta(x_t \mid x_{<t})$ is the model's predicted probability for that amino acid given the preceding tokens, and $m_t$ is a mask equal to 1 for positions in the variable fragment and 0 for prefix/suffix positions. This ensures that the model learns the sequence patterns relevant to functional variation while ignoring positions that are identical across all sequences. Fine-tuning was performed with AdamW optimization for 5 epochs, a batch size of 4, a learning rate of $1 \times 10^{-4}$, and gradient clipping of 1.0. A 10-fold cross-validation split was used for training and validation, with a 90/10 split of the sequences and a fixed random seed of 42. Generation temperature during training was set to 1.0.

### 4.5. Reinforcement learning

Following supervised fine-tuning, the model was further trained using a reinforcement learning framework that bias sequence generation toward variants predicted to be both viable and highly novel. During training, sequences were generated conditioned on the same N-terminal prefix used in fine-tuning, producing only the internal variable region, after which the C-terminal suffix was appended. Each generated sequence was assigned two scores: (i) a viability score, corresponding to the probability of being classified as viable by a ProtBERT-based classifier developed in Rodrigues *et al.* (2026) [31], and (ii) a novelty score computed as described above. The novelty score was expressed as a percentile relative to the distribution of distances observed in the fine-tuned generated sequences. To further incentivize exploration of rare, highly novel sequences, a power-tail bonus was applied when a sequence exceeded the maximum novelty observed in the fine-tuned set. The per-sequence reward was calculated as:

Eq. (2):

$$R_i = 1(v_i \geq \tau) \cdot v_i^\alpha \cdot d_i^\beta$$

where $R_i$ is the reward for sequence $i$, $v_i$ is the predicted viability, $d_i$ is the novelty score, $\tau$ is the viability threshold, and $\alpha$ and $\beta$ are power coefficients controlling the relative influence of viability and diversity.

Policy updates were performed using a KL-regularized objective relative to a frozen reference copy of the generator model. Rewards were standardized within each batch, and only the top fraction of sequences (top-q) was retained; negative advantages were clipped to zero. The training objective minimized a weighted token-level negative log-likelihood (NLL) over the generated fragment, combined with KL and entropy regularization. Per-sample loss (Eq. 3) and batch-level loss (Eq. 4) were defined as:

Eq. 3:

$$L_i = A_i \cdot \text{NLL}_i + \text{KL}_\beta \cdot \text{KL}_i$$

Eq. 4:

$$L = \text{mean}_i(L_i) - \text{entropy\_bonus} \cdot H$$

where $A_i$ is the scaled advantage for sequence $i$, $\text{KL}_i$ is the KL divergence between the current policy and the frozen reference policy, and $H$ is the entropy term promoting exploration. This dual-branch reward structure allowed the model to systematically balance exploitation of functional constraints with exploration of high-novelty sequence variants.

### 4.6. Sequence generation

For sequence generation, the models were prompted with the N-terminal prefix and autoregressively generated the variable fragment, using the prefix as context to produce novel sequences within the mutational window. The length of the generated fragment was sampled uniformly between 28 and 43 residues, reflecting the range observed in the training dataset for the mutated window. After generation of the variable region, the suffix was appended to reconstruct the full-length sequence. Generation was performed in batches of 10000 sequences using a fixed random seed, and sequences were sampled at two different temperatures, 0.8 and 1.2, to evaluate the effect of sampling stochasticity on sequence diversity.

**Funding & Acknowledgements**

This work was supported by FCT - Fundação para a Ciência e Tecnologia, I.P. under the LASIGE Research Unit, ref. UID/00408/2025, DOI https://doi.org/10.54499/UID/00408/2025, and partially supported by project 41, HfPT: Health from Portugal, funded by the Portuguese Plano de Recuperação e Resiliência. It was also partially supported by the CancerScan project which received funding from the European Union's Horizon Europe Research and Innovation Action (EIC Pathfinder Open) under grant agreement No. 101186829. Views and opinions expressed are however those of the author(s) only and do not necessarily reflect those of the European Union or the European Innovation Council and SMEs Executive Agency. Neither the European Union nor the granting authority can be held responsible for them. Pedro Cotovio, and Lucas Ferraz acknowledge Fundação para a Ciência e a Tecnologia for the PhD grants, respectively, 2022.10557.BD and 2025.04034.BD.

**Data availability**

The dataset used for models training and fine-tuned was the viability set described by Bryant *et al.* (2021) [11] and is available at https://www.nature.com/articles/s41587-020-00793-4. Scripts and code used to generate the results in this study are publicly available at https://github.com/liseda-lab/genAAV.

**Declaration of Competing Interest**

The authors declare no competing financial interests or personal relationships that could have influenced the work presented in this paper.